\documentclass[twocolumn,aps,prc,showpacs,eqsecnum,floatfix,nofootinbib]{revtex4}

\usepackage{amssymb,amsmath}
\usepackage{epsfig}

\newcommand{\be}{\begin{equation}}
\newcommand{\ee}{\end{equation}}

\begin{document}

\title{The nuclear AC-Stark shift in super-intense laser fields} 

\author{Thomas~J.~\surname{B\"urvenich}}
\email{buervenich@fias.uni-frankfurt.de}
\affiliation{Max--Planck--Institut f\"ur Kernphysik,
Saupfercheckweg 1, 69117 Heidelberg, Germany}
\affiliation{Frankfurt Institute for Advanced Studies, Johann Wolfgang Goethe University,
Max-von-Laue-Str. 1,
60438 Frankfurt am Main,
Germany}

\author{J\"org~\surname{Evers}}
\email{joerg.evers@mpi-hd.mpg.de}

\author{Christoph~H.~\surname{Keitel}}
\email{keitel@mpi-hd.mpg.de}

\affiliation{Max--Planck--Institut f\"ur Kernphysik,
Saupfercheckweg 1, 69117 Heidelberg, Germany}

\date{\today}

\begin{abstract}
The direct interaction of super-intense laser fields in the optical frequency
domain with nuclei is studied. As main observable, we consider the
nuclear AC-Stark shift of low-lying nuclear states due to the off-resonant
excitation by the laser field. We include the case of accelerated nuclei
to be able to control the frequency and the intensity of the laser field 
in the nuclear rest frame over a wide range of parameters. We find 
that AC-Stark shifts of the same order as in typical quantum optical
 systems relative 
to the respective transition frequencies are feasible with 
state-of-the-art or near-future laser field intensities and moderate
acceleration of the target nuclei. Along with this shift, we find
laser-induced modifications to the proton root-mean-square radii and to the 
proton density distribution. We thus expect direct laser-nucleus
interaction to become of relevance together with other super-intense
light-matter interaction processes such as pair creation.
\end{abstract}

\maketitle

\section{Introduction}
In most branches of physics, a controlled manipulation
of the considered system has proven to be extremely useful to study 
fundamental system properties, and to facilitate a broad range of 
applications. A prominent example for this is quantum optics or laser 
physics in general~\cite{scully-zubairy,ficekbook,sargent},
for instance related to light-matter interactions on the level
of single quantum objects~\cite{single-objects,single-2}. 
Similar control is also possible at lower driving field 
frequencies, e.g., with NMR techniques in the microwave 
frequency region~\cite{microwave}.
Towards higher frequencies, in particular the development 
and deployment of high-intensity lasers 
have opened  the doors to new fascinating areas of physics of light-matter 
interactions. Laser fields reach and 
succeed the Coulomb field strength experienced by the electrons due 
to the nucleus and thus give rise to a plethora of exciting 
phenomena~\cite{review-mpi,review-rmp,ledingham02}. 

The above examples have in common that they focus on the interaction
of the driving fields with the outer electron shell of the atoms.
Regarding the interaction of strong laser fields with nuclei, however,
mostly  {\it indirect} reactions have been studied so far. In these reactions,
electrons or plasmas are encountered by a laser pulse and then, directly or 
by creating radiation, react with the 
nucleus.
Typical examples are the production of MeV X-rays in a plasma that is
generated by femtosecond laser pulses~\cite{schwoerer01}, the study
of $\gamma$-induced nuclear reactions in plasma
radiated by a super-intense laser~\cite{karsch98,leding2000}, or neutron 
production in laser plasma~\cite{pretzler98,izumi2002}.
Also, the coupling of nuclear and electronic transitions has been 
considered~\cite{el-nucl1}. Applications are lasing~\cite{el-nucl2},
the control of  M\"ossbauer spectra~\cite{el-nucl3,el-nucl5},
or inversionless amplification~\cite{el-nucl4}.
Further applications include optically induced nuclear 
fission~\cite{boyer98} and fusion~\cite{Ditmire99,krainov05},
nuclear reactions, isomer excitations~\cite{leding03},  
or nuclear collisions~\cite{solokov04}.

In contrast, {\it direct} laser-nucleus
interactions do not involve intermediating particles such as electrons 
or gamma ray photons. So far, however, such direct interactions have 
mostly been dismissed because of small interaction 
matrix elements~\cite{matinyan98}. Rare exceptions
study direct laser- and x-ray-nucleus interactions 
in the context of $\beta$ decay~\cite{becker83} 
or x-ray-driven gamma emission of nuclei \cite{carroll01}.
On the other hand, effects such as laser-induced pair creation
which previously had been neglected for the same reason of small
interaction matrix elements, are now being studied, see for 
example~\cite{burke97,pair4,pair5,pair6,mueller03,pair1,pair2}. 
The reason is that it can be expected that present and upcoming technology 
will allow to  enter regimes where these traditionally neglected processes 
become possible. 

In~\cite{letter}, we have shown that direct laser-nucleus interactions 
may indeed become of relevance in future experiments employing x-ray lasers, 
opening the field of nuclear quantum optics. 
In particular, the coherence of the laser light expected from 
new sources such as TESLA XFEL~\cite{tesla} is the 
essential feature which may allow to access extended coherence 
or interference phenomena reminiscent of atomic quantum optics.
Such laser facilities, especially in conjunction with moderate 
acceleration of the target nuclei to match photon and transition frequency, 
may thus enable one to achieve nuclear Rabi oscillations, photon echoes or
more advanced quantum optical schemes~\cite{scully-zubairy} in nuclei.

This in principle may allow for a considerable range of applications:
As ultimate goal, one may hope that strong laser fields could be 
utilized as tools for preparation, control 
and detection methods in nuclear physics.  Possible applications
could be the control of the reaction channels in laser-nucleus 
interactions, i.e., switch between pair creation, nuclear excitation, 
fragmentation, fission or other processes. Furthermore, and based on the 
experience of high-precision laser spectroscopy for atomic and 
molecular systems, lasers might be employed to  measure high-resolution 
spectra especially of low-lying nuclear states, as well as nuclear 
properties such as energies, 
lifetimes, and transition moments. Laser-assisted preparation of nuclear 
states may also serve to find new effects or reaction channels in nuclear 
reactions. In addition, some observables may allow to measure
properties of nuclei such as transition dipole moments and transition energies
independent of nuclear models~\cite{letter}.

From a comparison with atomic physics, it appears obvious that a near-resonant 
driving of nuclear transitions as studied in~\cite{letter} is the most 
promising approach to laser-nucleus interactions. The large 
transition frequencies in nuclei, however, make this challenging,
and require high-frequency laser facilities, possibly assisted by an 
acceleration of the target nuclei. Such coherent high-frequency light
sources, however, are rare as compared to corresponding light sources
at optical frequencies.
Thus the question arises, whether
direct laser-nucleus interactions are also possible and of relevance
with super-intense
laser fields in the optical frequency region, far off resonance with
the considered nuclear transitions. The obvious advantage of this 
approach is a relaxation of the demands on the facilitated laser 
source with respect to frequency.

Therefore, in this study, we investigate AC-Stark shifts
 of single-particle proton states
 in the presence of off-resonant super-intense laser fields. 
We find that these shifts may serve as a signature of direct laser-nucleus 
interactions. In the lab frame, the considered laser fields are in the 
optical frequency region (${\cal O}(1$~eV)). Head-on collisions 
of the laser field with accelerated nuclei allow to increase the frequency
and the intensity of the photons in the rest frame of the nuclei.
The required nuclear properties are calculated with the help of a
relativistic mean-field model. Relativistic mean-field models, and more
generally self-consistent mean-field models, provide a wealth of information
on the nuclear ground state in a converged calculation, such as 
the binding energy, the proton, neutron, and charge density, as well
as all single-particle wave-functions. The latter are of most relevance for
the present study.

We find that with the help of a moderate acceleration of the 
target nuclei, present and near-future super-intense laser fields
may induce AC-Stark shifts which relative to the respective transition
frequencies are of similar order as found in typical quantum
optical setups. 
Our primary observable, the AC-Stark shift, is closely related
to work in atomic physics in order to facilitate a comparison
of these two branches. It should be noted, however, that while 
the nuclear AC-Stark shift is closely related  to the atomic counterpart, 
there are some interesting differences, which may allow
for physical processes exclusively available in nuclei. These
differences will briefly be discussed in the final part.
As a first step in this direction, we further study
proton root-mean-square (rms) radii and proton densities under influence 
of off-resonant 
super-intense laser fields as typical observables in nuclear physics.

The article is structured as follows:
In Sec.~\ref{framework}, we describe the laser-nucleus interaction 
employed in this study as well as the nuclear model that is used to 
calculate the single-particle wave-functions. We discuss the 
computational procedure and present possible observables.  
Section~\ref{results} presents the numerical results of the AC-Stark 
shift calculations and discusses their implications. We also relate them to 
the atomic case. Section~\ref{conclusions} discusses and summarizes
the results.

\section{\label{framework}Theoretical framework}
%
\subsection{\label{sec-int}Laser-nucleus interaction}
We treat the laser-nucleus interaction in the electric dipole 
approximation, in which the (non-relativistic) interaction term 
in the length gauge is given by~\cite{sargent,eisenberg}
\begin{equation}
H_I = - e \vec{E}(t) \cdot \vec{r}\,.
\end{equation}
Here, $e=|e|$ is the electron charge,
$\vec{E}(t)$ is the electric field, and $\vec{r}$ the position operator.
For light linearly polarized in $z$-direction, this reduces to 
$H_I = - e E(t)z$.
The total Hamiltonian of our system is thus
\begin{equation}
H = H_0 + H_I \,,
\end{equation}
where $H_0$ denotes the nuclear Hamilton operator that is specified 
by the nuclear model employed and will be described in
Section~\ref{nuclear_model}.
A few comments on the choice of this interaction are in order. The 
spatial dependence of the electric laser field
is neglected in the dipole approximation due to the small extension 
of the nucleus of the order of a few Fermi. 
The magnetic fields can be neglected due to the smallness of the 
laser-nucleus interaction. Here we have an important difference to 
atomic systems: Intensities considered large on atomic scales
(they compete with the Coulomb force of the nucleus), typically 
are still weak as compared to the much stronger force between nucleons.
Thus a non-relativistic treatment is justified in our case. The nuclear model
employed in this study provides a covariant framework for the nuclear
ground-state description. Note, however, that the nucleons within the 
nucleus move non-relativistically. The predominant relativistic feature
is the strong spin-orbit force in nuclei, which is an intrinsic ingredient in
a covariant description employing strong scalar and vector fields.

In axial symmetry, the proton single-particle wave-functions can be written 
as~\cite{rutz}
\begin{align}
\psi_i&(z, \rho,\phi) = \psi_i^{\eta\sigma}(z,\rho,\phi)  
\nonumber \\
&=\left(\begin{array}{c}
\phi^{++}_{\eta_i m_i \pi_i}(z,\rho) 
     \exp\left[ i(m_i - \frac{1}{2})\phi\right] \\[2ex]
\phi^{+-}_{\eta_i m_i \pi_i}(z,\rho) 
     \exp\left[ i(m_i + \frac{1}{2})\phi\right] \\[2ex]
\phi^{-+}_{\eta_i m_i \pi_i}(z,\rho) 
     \exp\left[ i(m_i - \frac{1}{2})\phi\right] \\[2ex]
\phi^{--}_{\eta_i m_i \pi_i}(z,\rho) 
     \exp\left[ i(m_i + \frac{1}{2})\phi\right] 
\end{array}\right) \,,
\end{align}
where $n_i, m_i, \pi_i$ are radial quantum number, the projection of the 
total angular momentum on the symmetry ($z-$) axis, and the parity.
The corresponding eigenvalue equations for $m_i$ and $\pi_i$ read
$\hat{J}_z \psi_i = m_i \psi_i$
and
$\hat{P} \psi_i = \pi_i \psi_i$.
The overlap of two states in axial symmetry is
\begin{align}
\langle\psi_i|\psi_j\rangle =& \delta_{m_i m_j}\delta_{\pi_i\pi_j} 2\pi
\int_{-\infty}^{\infty} dz
\int_{0}^\infty \rho d\rho 
\nonumber \\
&\times\sum_{\eta,\sigma} \phi^{\eta\sigma}_{\eta_i m_i \pi_i}(z,\rho)
\phi^{\eta\sigma}_{\eta_j m_j \pi_j}(z,\rho) \,.
\end{align}
The condition $m_i = m_j$ follows from 
$\int_0^{2\pi} d\phi\: e^{i\,\Delta m \,\phi} = 0$  for $\Delta m \neq 0$.
The interaction Hamiltonian $H_I$ introduces no additional $\phi$-dependence and
the operator $z$ commutes with $J_z$,
hence the condition $m_i = m_j$ persists.
Furthermore, since  it is an odd function of $z$, we now have for 
non-vanishing matrix elements the condition $\pi_i \neq \pi_j$. The
matrix elements read
\begin{align}
\langle\psi_i|&z|\psi_j\rangle = \delta_{m_i m_j}\delta_{\pi_i[\pi_j\times(-1)]}
2\pi \int_{-\infty}^{\infty} dz
\int_{0}^\infty \rho d\rho 
\nonumber \\
&\times \sum_{\eta,\sigma} \phi^{\eta\sigma}_{\eta_i m_i \pi_i}(z,\rho)
\times z \times \phi^{\eta\sigma}_{\eta_j m_j \pi_j}(z,\rho) \,.
\end{align}
%

\subsection{Observables}
We focus on two observables relevant to nuclei exposed to 
super-intense laser fields. First, the proton energy shifts 
themselves are --in principle-- observable. Second, if the AC 
Stark shifts are large and the single-particle states are affected 
to a certain extent, the nuclear density experiences changes,
hence density or form-factor related observables become of interest.

\subsubsection{Stark shift}
The AC Stark shifts of the proton single-particle states in the 
laser field can be calculated equivalently to the case of electron 
states in the atom. A semi-classical calculation of the dynamic 
Stark shift yields~\cite{sakurai,haas}
\begin{equation}
\Delta E_n = \frac{1}{4} \sum_{m, \pm} 
\frac{\langle n| H_I | m\rangle \langle m| H_I | n\rangle}
{\epsilon_n - \epsilon_m \pm \hbar \nu + i\hbar \epsilon} \,.
\label{stark-shift}
\end{equation}
The (unperturbed) single-particle energies are denoted by $\epsilon_m$. 
Note that $\Delta E$ arises as a second-order perturbation effect since 
the single-particle wave-functions have good parity (this is also true 
in the atomic case). As discussed in~\cite{haas}, the 
quantum-mechanical calculation yields the same result in the limit of 
large photon number, which applies to our study.

In the limit $\hbar\nu \ll \Delta \epsilon = \epsilon_n-\epsilon_m$, 
i.e., for laser field photon energies well below the nuclear transition
frequencies, the laser-frequency dependence in the denominator
drops out, leaving us with
\begin{equation}
\Delta E_n^{\ll} = \frac{1}{2} \sum_{m \ne n} \frac{\langle n| 
H_I | m\rangle \langle m| H_I | n\rangle}{\epsilon_n-\epsilon_m} \,.
\label{stark-adiab}
\end{equation}
The same expression is obtained in a time-averaged calculation in 
the adiabatic limit~\cite{typ96}. In the following, we use 
expression Eq.~(\ref{stark-adiab}) to quantify the Stark shifts, 
since we focus on the off-resonant excitation of the nuclear
transitions, such that $\hbar\nu \ll \Delta \epsilon$ is fulfilled 
in all cases considered.

\subsubsection{Density-related observables}
\label{density-observables}
The actual proton density of the nucleus exposed to the laser 
field can be computed by taking into account perturbatively the 
corrections to the wave-functions due to the interaction with 
the laser field through $H_I$.
We write the spatial part  $|\phi_n\rangle$ of the total wave function 
in second-order perturbation theory as
\be
|\phi_n\rangle = |n^0\rangle + |n^1\rangle + |n^2\rangle \,,
\ee
where the superscript indicates the order of perturbation.
In the adiabatic limit, one obtains~\cite{typ96}
\begin{align}
|\phi_n\rangle = |n^0\rangle  &+ \sum_{k} a^1_{kn}\sin(\omega_Lt) 
|k\rangle 
\nonumber \\
&+ \sum_k a^2_{kn} \sin^2(\omega_L t)|k\rangle \,.
\end{align}
The first ($a^1_{kn}$) and second ($a^2_{kn}$) order expansion 
coefficients read~\cite{typ96}
\begin{subequations}
\label{coefficients}
\begin{align}
a^1_{nn} =& \langle n^0| n^1\rangle = 0 \,,  \\[1ex]
a^1_{kn} =& \langle k^0| n^1\rangle = \frac{\langle k|H_I|n\rangle}
{E^0_n-E^0_k} \qquad (k\ne n) \,, 
\allowdisplaybreaks[2] \\[1ex]
a^2_{nn} =&  \langle n^0 | n^2\rangle = - \frac{1}{2} \sum_{m \ne n}
\frac{|\langle m|H_I | n \rangle|^2}{(E^0_n-E^0_m)^2}\,,  
\allowdisplaybreaks[2] \\[1ex]
a^2_{kn} =& \langle k^0|n^2\rangle = \sum_{m \ne n} 
\frac{\langle k| H_I| m\rangle  \langle
m|H_I|n\rangle}{(E^0_n-E^0_k)(E^0_n-E^0_m)} 
\nonumber \\
&- \underbrace{\frac{\langle k|H_I|n\rangle \langle n|H_I|
n\rangle}{(E^0_n-E^0_k)^2}}_{=0} \qquad (k\ne n) \,. \label{coeff-d}
\end{align}
\end{subequations}
The last addend of Eq.~(\ref{coeff-d}) vanishes from parity.
In this study we are interested in the time-averaged single-particle 
densities from which we can compute the proton rms radius and the 
proton quadrupole moment. Using 
\begin{subequations}
\begin{align} 
\lim_{T\rightarrow\infty} \frac{1}{T}\int_0^T dt \sin(ct) =&0 \,, 
\allowdisplaybreaks[2] \\
\lim_{T\rightarrow\infty} \frac{1}{T} \int_0^T dt \sin^2(ct) =& 
 \frac{1}{2}\,, 
 \allowdisplaybreaks[2] \\
\lim_{T\rightarrow\infty} \frac{1}{T}\int_0^T dt \sin^3(ct) =&0 \,,
\allowdisplaybreaks[2] \\
\lim_{T\rightarrow\infty} \frac{1}{T}\int_0^T dt \sin^4(ct) =& 3/8 \,, 
\end{align}
\end{subequations}
we can compute the average single-particle density in coordinate 
space as~\cite{ring-schuck}
\be 
\overline{\rho}_l = \overline{\phi_l^*\phi_l} \ne
\overline{\phi^*_l}\times\overline{\phi_l} \,,
\ee
and obtain
\begin{align}
\overline{\rho}_l =& \phi^{0*}_l\phi^0_l 
+ \sum_i a^2_{li} \phi^{0*}_l\phi_i 
\nonumber \\
&+ \sum_{i,j} \Big(\frac{3}{8} a^2_{li}a^2_{lj} + \frac{1}{2}
a^1_{li}a^1_{lj}\Big) \phi^{0*}_i\phi^0_j \,.
\end{align}
This density is used for the calculation of the ground-state proton radius 
and deformation as shown below. These are standard observables used to
calibrate and judge the predictive power of nuclear models using known
experimental data~\cite{adjustments}.

The rms radius of the proton density is defined as~\cite{ring-schuck}
\be
r_{rms}^p = \sqrt{\frac{\int d^3x r^2 \rho^p(\vec{x})}
{\int d^3x \rho^p(\vec{x})}} \,,
\label{rms}
\ee
with $r = \sqrt{x^2+y^2+z^2}$, $\rho^p(\vec{x})$ is the proton point 
density. Note that this definition also holds for non-spherical
density distributions. The rms radius is related to the spatial extension 
of the density distribution. The experimentally accessible quantity in 
nuclei is the nuclear charge radius which can be extracted from the
corresponding measured form factor.

The spherical quadrupole moment in axial symmetry reads~\cite{ring-schuck}
\be
Q_{20} = \frac{1}{2} \sqrt{\frac{5}{4\pi}} \int d^3x \rho^p(\vec{x}) 
(2z^2-r^2)\,,
\label{q2}
\ee
where $r = \sqrt{x^2+y^2}$. Positive values of $Q_{20}$ denote 
cigar-like shapes, while negative values correspond to disk-like 
nuclear density distributions. Since the quadrupole moment integrates 
over the proton density it shows a mass dependence.
A dimensionless quantity without such mass dependence is given by
\be
\beta_2 \equiv \beta_{20} = \frac{4\pi}{3nR^2}Q_{20}\,,
\ee
where $n = \int d^3x \rho^p$, and $R = 1.2 ~{\rm fm} \times A^{1/3}$ 
is an approximation of the nuclear radius (A is the total number 
of nucleons).

\subsection{\label{nuclear_model}The nuclear model}
A quantitative estimate of the nuclear dynamic Stark shifts 
demands realistic proton single-particle wave functions which we 
obtain by employing the relativistic mean-field (RMF) model for the 
ground-state calculation of the nucleus. 
Though we have no guaranty that these wave-functions yield a close 
approximation to nature, the success of the RMF approach supports 
our choice~\cite{rei89,furn04}. Moreover, these wave functions
do not suffer from known deficiencies of other approaches, e.g., 
the wrong asymptotics of wave functions obtained in a harmonic 
oscillator potential. 

The RMF model~\cite{serot86,rei89,bender-review,ring-rmf} has 
historically been designed as a renormalizable meson-field theory
for nuclear matter and finite nuclei. The realization of nonlinear
self-interactions of the scalar meson led to a quantitative description 
of nuclear ground states.
As a self-consistent mean-field model (for a comprehensive review 
see~\cite{bender-review}), its ansatz is a Lagrangian or Hamiltonian
that incorporates the effective, in-medium nucleon-nucleon interaction. 
In contrast to macroscopic-microscopic approaches, no assumptions on 
the nuclear potential or density are made. RMF models yield the binding energy
and all single-particle wave-functions in one calculation, from which several
other kinds of observables can be obtained.

Recently, self-consistent models have undergone a 
reinterpretation~\cite{furn04} which explains their quantitative
success in view of the facts that nucleons are composite objects and 
that the mesons employed in RMF have only a lose correspondence to 
the physical meson spectrum \cite{foot-meson}
They are seen as covariant {\em Kohn-Sham schemes}~\cite{dreiz90}  
and as approximations to the true functional of the nuclear ground 
state. According to the {\em Hohenberg-Kohn theorem}, the exact 
ground-state functional does exist. However, this theorem does not 
provide a handle to construct it (it is non-constructive).
As a Kohn-Sham scheme, the RMF model can incorporate certain 
ground-state correlations and yields a ground-state description 
beyond the {\em literal} mean-field picture. RMF models  are 
effective field theories for nuclei below an energy scale
of $\Lambda \approx 1$~GeV, separating the long- and intermediate-range 
nuclear physics from short-distance physics, involving, i.e., 
short-range correlations, nucleon form factors, vacuum polarization etc, 
which is absorbed into the various terms and coupling constants.

The strong attractive scalar ($S \approx -400~{\rm MeV}$) and repulsive 
vector ($V \approx +350~{\rm MeV}$) fields provide both the binding 
mechanism ($S + V \approx -50~{\rm MeV}$) and the strong spin-orbit force  
($S-V \approx -750~{\rm MeV}$) of both right sign and magnitude.

The RMF model is based on phenomenology and needs an adjustment of 
its (phenomenologically introduced) coupling constants~\cite{adjustments}.  
We have chosen the parameterization NL3~\cite{lalazissis}, which is among 
the most successful parameterizations available.

\subsection{\label{sec-comp}Computational procedure}
The stationary mean-field equations are solved with a {\tt C++} code 
on a grid in coordinate space in axial symmetry. The wave functions 
are written out and then processed to compute the dipole matrix 
elements $\langle a|z|b\rangle$ between respective proton states 
$a$ and $b$. Matrix manipulations and integrations are done using 
{\tt Python} together with the modules {\tt NumArray}~\cite{numarray} 
and {\tt SciPy}~\cite{scipy}. We have neglected pairing in our 
mean-field calculation to be consistent with the following
computation of the matrix elements and the Stark shifts. While 
pairing is important for a highly accurate description of ground-state 
energy and deformation, it is not relevant for our purpose.

Uncertainties in the calculations of the dipole matrix 
elements stem from the calculated radial components of the wave functions. 
Still, mean-field wave-functions can be considered realistic for 
nuclei close to the valley of stability and for well-bound states, which we
consider here. Thus we will not reach the accuracy reached in QED 
calculations, but we still can expect solid quantitative 
predictions of the AC Stark shifts. This justifies further approximations 
that introduce uncertainties within this framework. 
We have neglected the influence of the center of mass motion of the 
nucleus and the (very weak) coupling of the neutrons to the laser, thus
no effective charges were introduced in our calculations. Furthermore, 
all effects beyond the electric dipole approximation, which are related 
to the magnetic field contributions of the laser field 
have been omitted. In this respect, one should note that it is not the
laser field intensity itself, but rather the effective coupling 
$H_I$ relative to the mean-field Hamiltonian $H_0$ which determines 
whether the electromagnetic field induces large perturbation on the 
nuclear ground state or not. Thus approximations are valid in the 
nuclear case for intensities where the same approximations break down 
for the calculations of atomic or molecular systems. 

In this work we assume that a spheroidal nucleus will have its
symmetry axis aligned with the direction of the laser field. This way, 
the interaction with the laser field does not destroy axial symmetry of 
the system which we employ in our numerical solution of the mean-field 
equations. Since most spheroidal nuclei possess no static dipole moment, this 
alignment will not take place naturally. However, alignment
can take place under the following conditions: a) the nucleus has a
reflection-asymmetric ground-state shape and thus a static electric 
dipole moment (there are rare cases)~\cite{butler}; 
b) we employ an additional electric field gradient in the polarization 
direction of the laser field. The interaction with the quadrupole moment 
of the nucleus then leads to alignment since the interaction energy of 
the quadrupole moment with the external static electric field is given 
by $W = -\frac{1}{4} \Big( \frac{\partial E_z}{\partial z}\Big) e Q_z$
assuming that the electric field is pointing in $z$ direction. Without such 
an alignment, the nucleus exposed to the laser field will experience 
shape changes that lead to triaxial shapes
and the dipole matrix elements will slightly differ.
The size of the Stark shifts, however, will be similar to the ones 
calculated in the aligned case.

\subsection{Laser-Nucleus collisions}
Both laser frequency and intensity in the nuclear rest frame can 
be effectively increased by letting the nucleus and laser collide 
head-on. In the rest frame of the nucleus, the  Doppler shifted 
electric field strength $E_N$ and the frequency $\nu_N$ are given by
\begin{eqnarray}
E_N &=& \sqrt{\frac{1+\beta}{1-\beta}} E_L = (1+\beta)\gamma E_L \,, \\
\nu_N &=& \sqrt{\frac{1+\beta}{1-\beta}} \nu_L = (1+\beta)\gamma \nu_L \,,
\end{eqnarray}
where subscript $N$ denotes the nuclear rest frame and $L$ the 
laboratory frame, respectively.

\begin{figure*}[t]
\centerline{\epsfxsize=16cm \epsfbox{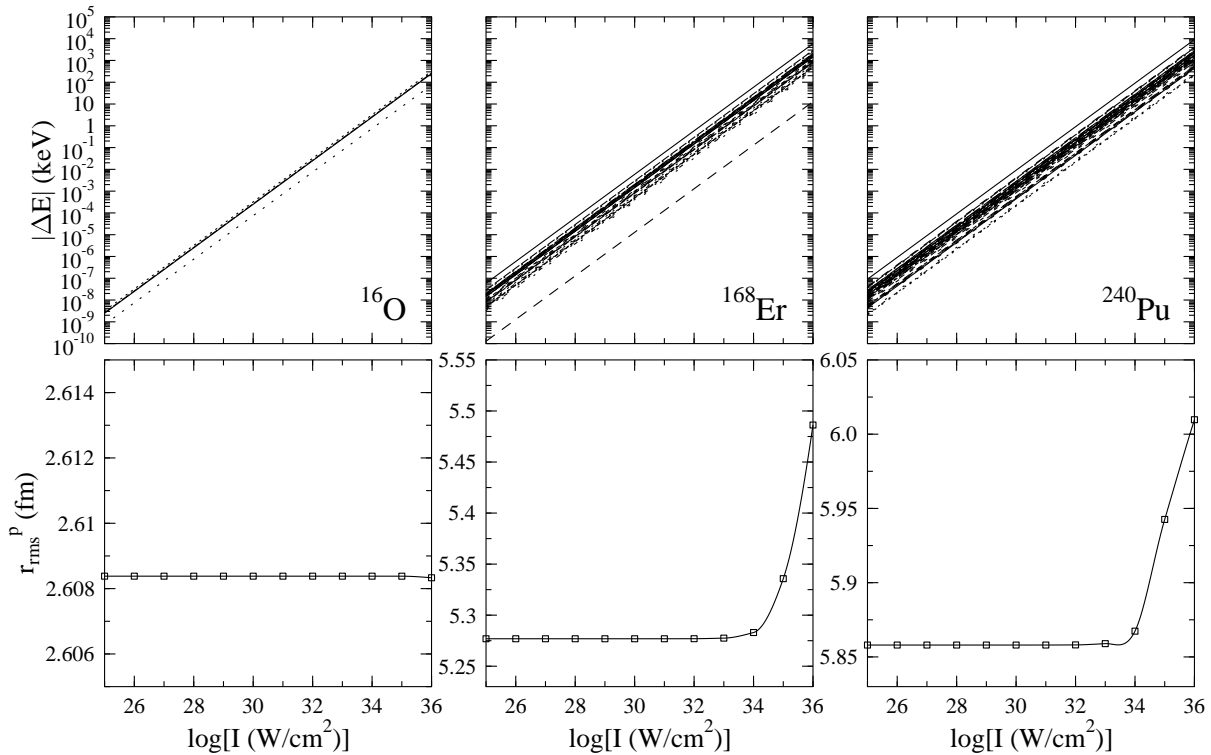}}
\caption{AC-Stark shifts of the proton single-particle states in the 
laser field (upper row) as a function
of laser intensity in the nuclear rest frame for the nuclei as indicated, 
and proton rms radii (lower row) as a function of laser intensity in the 
nuclear rest frame. Each line in the upper figures corresponds to a Stark 
shift of a proton single-particle level. The widths of these bands 
characterize the spread in these shifts. In the lower figure, the
square dots indicate the calculated results, which for convenience 
are connected by the thin line.}
\label{all}
\end{figure*}

For $\gamma$ factors of about $1000~(4000)$, on has $\beta \approx 1~(1)$, 
and we obtain $(1+\beta)\gamma \approx 2000~(8000)$. Since the laser 
intensity is proportional to $E^2$, we find amplification by a factor of
$4.0 \times 10^6 ~(6.4 \times 10^7)$. Assuming lab-frame intensities of 
$I_L \approx 10^{22-24}~{\rm W/cm}^2$, we can reach $I_N 
\approx 10^{28-31}~{\rm W/cm^2}$. Intensities of 
$I_L \approx 10^{28}~{\rm W/cm}^2$ are in reach in the near future, 
and the higher the laser intensity, the smaller is the necessary $\gamma$
factor of the accelerated nuclei. As will be shown below, in order
to reach AC Stark shifts comparable to typical shifts in atomic systems
with respect to the transition frequencies, only moderate $\gamma$ shifts
are necessary.
For these cases of $\gamma \approx 1000-4000$, employing optical lasers with $E \approx 1~{\rm eV}$, 
photon energies of about $2-10~{\rm keV}$ result, which is still smaller 
than typical energy differences of proton
single-particles energies of a few MeV (deeply bound states) or some 
hundreds of keV (near the Fermi edge). Choosing IR-lasers in the first 
place yields even smaller photon energies.

In the following, we discuss frequencies and intensities in the 
nuclear rest frame. Note that it is not important whether 
the assumed values are reached via a large velocity as compared
to the lab frame, a powerful laser facility, or
a combination of both.
Both high-intensity laser as well as ion accelerators are available today
or in the near future, albeit mostly in separate places. A promising
ansatz to reach experimental conditions as required for the physics discussed
here would be to install and combine both types of facilities in one laboratory.

\section{\label{results}Results}
Figure~\ref{all} shows the AC Stark shifts of single proton states 
in the nuclei $^{16}$O, $^{168}$Er, and $^{240}$Pu, as well as the 
corresponding rms proton radii. We have chosen these nuclei as typical
representatives of light, intermediate, and heavy nuclei. Their lowest 
measured E1 excitations lie at 7.117 MeV ($^{16}$O), 1.359 MeV ($^{168}$Er),
and  0.555 MeV ($^{240}$Pu), respectively~\cite{toi}. Thus, transitions will not 
be excited by the considered laser energies of ${\cal O({\rm keV})}$ in 
the nuclear 
rest frame. Lower excitations of even parity would require two- or higher-order 
photon processes, and their energies are still
more than 20 keV above the ground state energy. Hence we can treat the
Stark effect separately from nuclear excitation mechanisms.

Shifts of $\approx 1$~keV are reached at intensities of
roughly $10^{34}$ W/cm$^2$ for oxygen, and  $10^{32}$ W/cm$^2$ for the 
heavier systems. As discussed above, these shifts are approximately a factor 
of 10-1000 smaller than typical energy differences of single-particle 
levels close to the Fermi edge. As expected, in absolute terms,
they are much larger than shifts appearing in atomic systems, but
may also surpass them in relative terms, see the 
end of this section for details.
The size of the shifts depends both on the matrix elements 
$\langle m |H_I | n\rangle$ as well as on  the number of states 
contributing with dipole moments and their corresponding 
single-particle energies.
Since oxygen has only 8 protons, the effects are rather small. 
There are no significant changes in the proton rms radius.
The Stark shifts in $^{168}$Er, and $^{240}$Pu are (on the average) 
larger than in oxygen due
to the increased number of states. Also, changes in the proton radii 
set in above $I=10^{33}$ W/cm$^2$.

For the medium and heavy nuclei under consideration, shape changes 
of the nucleus (as reflected in the increasing rms radii) also lead 
to an increase of the quadrupole moment of these systems, see 
Section~\ref{density-observables}, changing
the moments of inertia. This will in turn alter the rotational excited 
states of these systems.

\begin{figure*}[t]
\begin{minipage}[t]{80mm}
\centerline{\epsfxsize=8cm \epsfbox{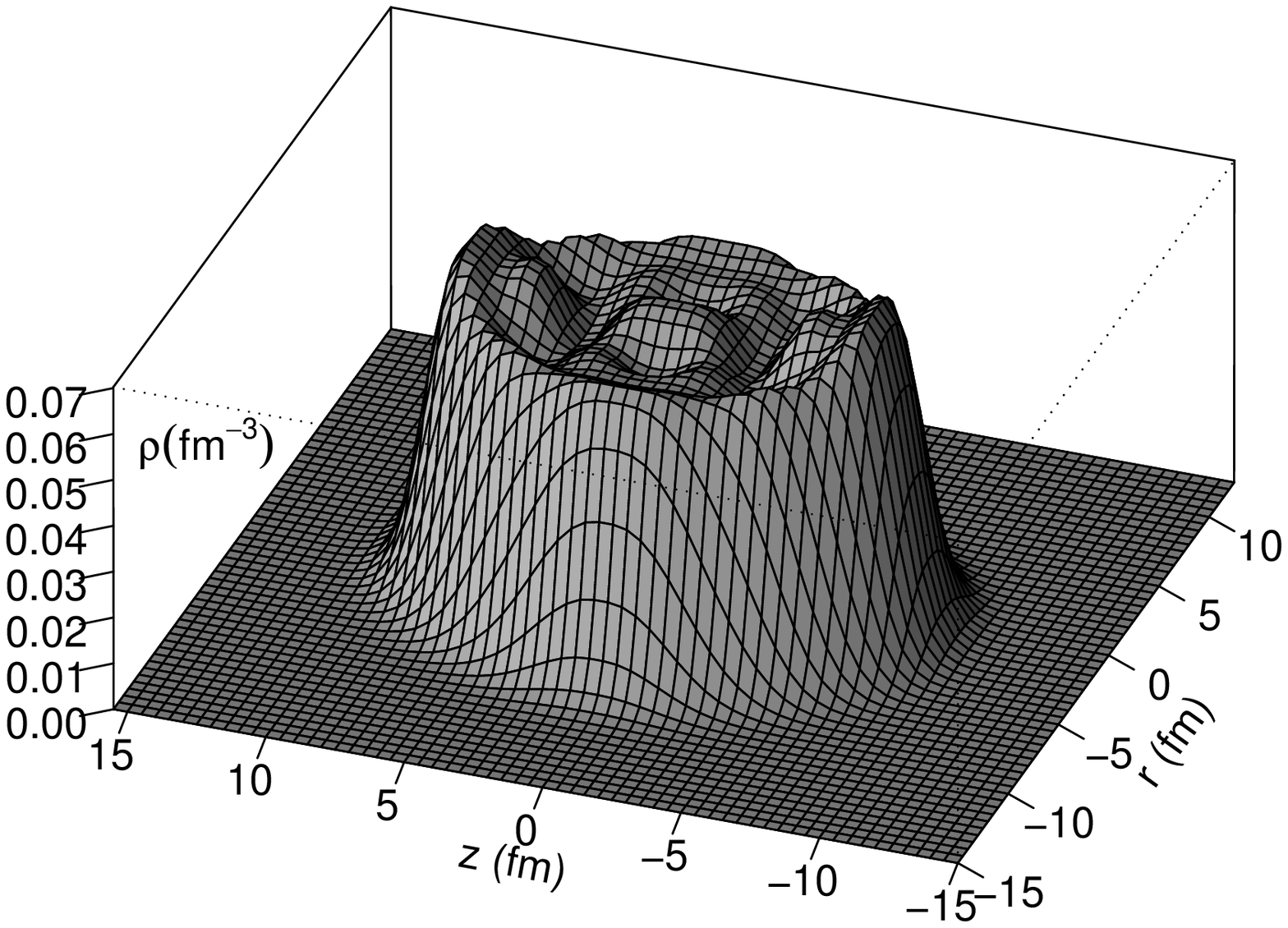}}
\end{minipage}
\begin{minipage}[t]{80mm}
\centerline{\epsfxsize=8cm \epsfbox{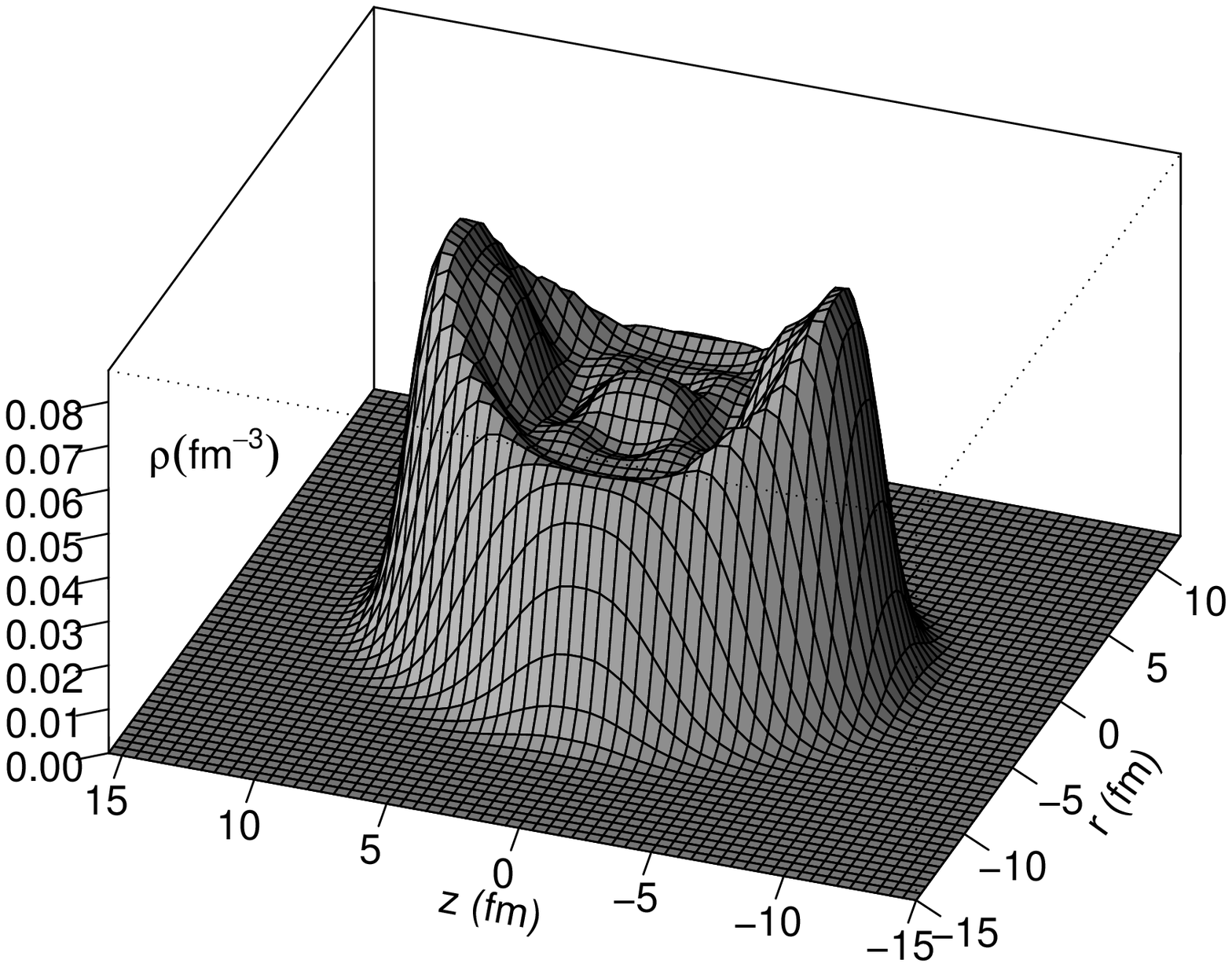}}
\end{minipage}
\caption{Proton density of $^{240}$Pu for intensities of 
$I = 10^{25}$~W/cm$^2$ (left) and $I = 10^{35}$~W/cm$^2$ (right).}
\label{pu240_dens}
\end{figure*}

Figure~\ref{pu240_dens} displays the proton density for laser intensities 
of $I = 10^{25}$~W/cm$^2$ (left) and $I = 10^{35}$~W/cm$^2$ (right). At
$10^{25}$~W/cm$^2$, the density profile resembles the ground-state
density, no differences are visible. This is due to the fact that 
Stark shifts and, correspondingly, changes of the single-particle
wave-functions are small. At the higher intensity, Stark shifts reach values of 
a few hundred keV. This is certainly the limit of our adiabaticity 
assumption. The proton density is slightly
extended in $z$- and $r$-directions. Even more prominently, the density 
close to the center of the nucleus is reduced, and the {\em poles} of 
the proton density get enhanced. This might be related to the fact 
that the dipole matrix elements yield largest contributions for states 
close to the Fermi edge, where high total angular momentum projections 
on the $z$-axis occur. These states are localized at large $z$-values 
and thus lead to the visible enhancement. This rearrangement of
the nucleus leads to an increase of the quadrupole moment, see Eq. (\ref{q2}).
Significant changes in this observable correspond to the respective changes in
the rms radii that are shown in Fig. \ref{all}.

We would like to classify the various processes taking place for a 
nucleus in a super-intense laser-field according to the nuclear 
rest-frame laser intensity (see 
Ref.~\cite{review-mpi,review-rmp,ledingham02,applications} for a discussion 
of effects relevant to laser-nucleus physics). The following hierarchy 
of effects can be constructed, going from low to high laser intensities in
the nuclear rest frame:
\\-
$I < 10^{29} $~W/cm$^2$: a) radiation from scattering off of 
the nucleus in the laser field; b) radiation from electrons surrounding 
the nucleus if the ion is not fully stripped; c) the AC Stark shifts are 
already comparable to the typical atomic shifts in relation to the 
transitional energies of single-particle states
\\-
$I \approx 10^{29} $~W/cm$^2$: This is the critical field 
strength~\cite{sauter31,schwinger51,brezin70} at which $e^+ e^-$ pair 
creation sets in~\cite{burke97,bamber99,mueller03}, additionally radiation 
is generated by created electrons or positrons that oscillate for a few 
cycles within the laser field
\\-
$I \ge 10^{32} $~W/cm$^2$: Direct laser-nuclear interactions 
become non-negligible, AC Stark shifts of proton states lead to a 
slight structural change of the nucleus, weak quadrupole oscillations take
place in the laser field; very weak quadrupole radiation sets in. 

We can compare the nuclear AC Stark effects with similar situations for 
atomic systems.
In typical non-resonant laser-atom systems which aim 
at measuring the AC Stark shifts in moderate laser fields, 
the relation of the energy shifts due 
to the laser fields compared to typical energy differences of 
${\cal O}({\rm eV})$ is $\approx 10^{-12}-10^{-10}$~\cite{beauvoir,haeffner03}.
In the nuclear case with energy differences of ${\cal O}({\rm MeV})$, 
this would correspond to AC Stark shifts of the order of $10^{-9}-10^{-7}$~keV,
as found in the low intensity regime of Fig.~\ref{all}. The corresponding 
intensities of $I = 10^{25}-10^{27}$~W/cm$^2$ are close to intensities that can be 
presently reached or are envisaged in the near future. Thus, nuclear AC Stark 
shifts that are similarly related 
to the typical transition energies as in the atomic case can in the future
be expected even without a pre-acceleration of the nuclei. It remains to be seen 
if these Stark shifts can be directly measured. Such kind of 
measurements, however, would demonstrate the direct laser-nucleus 
interactions in a very concise way. We would like to point out that the 
framework of
our Stark-shift calculations 
is based on continuous wave lasers, while in most realistic 
situations (laser-nucleus collisions, or super-intense
lasers incident on a fixed target) laser pulses will be employed. 
There, the calculated Stark shift sizes correspond to the central region 
of the laser pulse. 

In addition to the structural changes, due to the incident field, the 
nucleus will also
experience an oscillating center-of-mass motion, resulting in dipole
radiation perpendicular to the direction of the laser electric field and 
the beam axis (we do not consider the drift 
motion in beam direction, consistent with the non-relativistic treatment
of the light-matter interaction as discussed in Secs.~\ref{sec-int}
and~\ref{sec-comp}).
The oscillation extent $\Delta x_{pol}$, which is twice the amplitude 
of oscillation in polarization direction, can be calculated classically for
a structureless particle of charge $q$~\cite{mocken} in an electromagnetic
field, yielding $\Delta x_{pol} = (2|q|E_N)/(m \omega_N^2)$.
We estimate $\Delta x_{pol}$ for the nucleus $^{168}$Er, using 
$|q|=68\:e, ~m \approx 168\: {\rm u}$, 
for $E_N = 1.0\times 10^{16}\: {\rm V/cm}$.
As photon energies, we choose $\hbar \omega = 1$~eV and 
$\hbar \omega = 1$~keV, respectively.
The resulting oscillation extent is $\Delta x_{pol} 
\approx 3\cdot 10^{10}$~fm for $\hbar \omega = 1$eV and 
$\Delta x_{pol} \approx 3\cdot 10^{4}$~fm
for $\hbar \omega = 1$~keV.
This should be compared to the size of the nucleus, which is on the order
of 6~fm. Note also that here the nucleus is assumed to be free of electrons,
which increases the charge-to-mass ratio as compared to the case
of a singly charged ionic core after a single electron ionization. This
enhances the response of the nucleus to the incident field.
The dipole-type nuclear center of mass motion yields radiation, with total
radiation power given classically by~\cite{jackson}
\begin{equation}
P = \frac{c^2Z_0k^4}{12\pi} |\vec{p}|^2\,,
\end{equation}
where $Z_0 = \sqrt{\mu_0/\epsilon_0}$ is the impedance of the vacuum, 
and $\vec{p} = \int \vec{r} \rho^p(\vec{r}) d^3r$ is the dipole moment, 
respectively. 
Assuming this radiation to be emitted by photons of energy 
$E= \hbar \omega$, we can semi-classically estimate the time needed 
for the emission of one photon by
\begin{equation}
P = \frac{W}{t} = \frac{\hbar\omega}{t} \quad \Leftrightarrow \quad t 
= \frac{\hbar\omega}{P}\,.
\end{equation}
If we equate $\Delta x_{pol}$ with the length entering the dipole moment, 
for $^{168}$Er we obtain $|\vec{p}| = 68\:e \times \Delta x_{pol}$.
For photon energies of $\hbar \omega = 1$~eV we estimate
$t_N = 2\times 10^{-21}$~s, for $\hbar \omega = 1$~keV
we obtain $t_N = 2\times 10^{-18}$~s.

These emission times have to be compared to typical laser pulse
durations of 1-100~fs, which is the duration over which the
required field strength can be maintained. The average amount
of signal photons then further depends on the repetition rate of the 
experimental setup.
The radiation generated from nuclear quadrupole shape oscillations 
is suppressed as compared to this dipole radiation due to a center-of-mass
motion, which is 
unfortunate, since quadrupole radiation is a unique signal for the 
above discussed structural changes of the nucleus. Its detection may 
become feasible, however, once the required laser intensities
in the laboratory frame become available with high repetition rate.
Also, by preparing
large ensembles of nuclei flying head-on into the laser beam, the number
of individual interactions may become large enough for the 
detection of quadrupole 
radiation. 

Finally, we would like to return back to the structural properties of
nuclei in the laser field in contrast to atomic systems. The neutrons in nuclei
are likely to adiabatically follow the periodic changes of the proton states.
Thus, not only the proton density, but also the mass density undergoes
(tiny) quadrupole oscillations. Furthermore, non-closed-shell nuclei 
are superfluid systems, where the short-range pairing correlations soften
the Fermi edge and allow angular momentum paired nucleon-pairs to scatter
into energetically higher-lying orbits. The presence of the laser field 
will also effect the continuum states in the nucleus and thus we may
expect that the pairing correlations will be altered. This then,
in turn, affects the moment of inertia for deformed nuclei.

\section{\label{conclusions}Conclusions and Outlook}

Our work is motivated by the hope that externally controllable 
super-intense electromagnetic fields could enhance preparation, control 
and detection methods in nuclear physics similar to the 
tremendous success of such control methods in atomic and molecular
physics.
We have shown that a combination of cutting-edge lasers 
and ion accelerators available today or in the near future 
opens a pathway for the study of direct laser-nuclear interactions. 
These interactions do not involve intermediate particles such as 
external electrons accelerated by the laser pulse.
The potential for such applications obviously will increase with
improving laser and accelerator facilities. 
Ultimatively, resonant laser-nucleus interactions can be expected to 
be the most promising candidate for a direct manipulation of nuclei by coherent light,
but these require high-frequency laser sources.
As discussed in this article, however, also off-resonant 
super-intense laser fields in the 
optical frequency region may induce signatures of direct 
laser-nuclei interactions.
An important task, of course, remains to find convenient observables 
or experimental setups where the direct laser-nucleus interactions
can be observed or even crucially influence the outcome
of a desired measurement.
Examples could be processes sensitive to resonance conditions
subject to laser-induces shifts, or the observation of weak
quadrupole radiation emitted due to oscillatory excitations
of the nuclei.

Our main observable, the AC-Stark shift, is closely related
to work in atomic physics in order to facilitate a comparison
of these two branches. 
Despite the similarities of the nuclear AC-Stark shift calculations 
to the atomic case, however, several differences between the atomic and the 
nuclear case should be noted. 
First, in nuclei, the electromagnetic force is not the strongest force present.
The electric field generated by the laser field is a perturbation 
on top of the inter-proton Coulomb force, being itself a perturbation on the strong
nucleon-nucleon force governing to a major extent the
structure of the nucleus. 
In atoms, however, the electromagnetic force between electrons and 
the nucleus (and between electrons and electrons) is governing
structure of the atom.
Second, nuclei do not possess a central Coulomb
potential. Furthermore, the Coulomb force between protons within the nucleus 
is much stronger than the inter-atomic Coulomb force between electrons and 
the nucleus because of the much smaller size of the nucleus. 
Third, nuclei possess rich structural properties, i.e., collective 
excitations (rotations, vibrations, giant resonances), as well as
single-particle excitations. Especially with respect to collectivity,
they resemble more molecules than atoms. 
Obviously, these differences are of especial interest since they
reflect the unique properties of nuclei as compared to atomic systems.
Thus for the future, an investigation focused on typical observables 
in nuclear physics, possibly based on the
the specific differences of atomic and nuclear physics,
is desirable. The proton rms radii discussed in this work are a 
first step in this direction, even though we found that a measurement 
of laser-induced changes in these radii is rather challenging.

Finally, our study could also be of relevance for situations
where {\it indirect} laser-nucleus-interactions occur, as it provides
data on the modification of nuclear properties under the 
influence of external fields. For example, a laser field 
utilized to create secondary particles which in turn interact
with nuclei can also be expected to modify the interaction of the 
nuclei with the auxiliary particles.

In summary, we have studied the direct off-resonant interaction of super-intense 
fields in the optical frequency region with nuclei.
In particular, laser fields which are in the optical region in the 
lab frame were considered, possibly in combination with an acceleration
of the target nuclei. Then already field intensities available
now or in the near future can induce AC-Stark shifts of the 
same order as in typical quantum optical systems relative to the 
respective transition frequencies.
We thus expect these direct laser-nucleus
interaction to become of relevance together with other super-intense
light-matter interaction processes such as pair creation.

\section*{Acknowledgments}
TJB is grateful to C. M\"uller and U. D. Jentschura for helpful comments.



\begin{thebibliography}{99}
\bibitem{foot-meson}
References to physical mesons, and the constraints
introduced into the Lagrangian, can be further reduced by using RMF 
models with point couplings (contact interactions) in the effective 
Lagrangian~\cite{nhm,rmf-pc,rmf-pc2}, which have a predictive 
power comparable to RMF models with auxiliary boson fields.
\end{thebibliography}
\end{document}